\documentclass{article}

\usepackage{pdfpages}

\usepackage{spconf,amsmath,graphicx}
\usepackage{blindtext}
\usepackage{hyperref}

\usepackage{amssymb,bbm}
\title{Continuous descriptor-based control for deep audio synthesis}
\name{Ninon Devis*, Nils Demerlé*, Sarah Nabi*, David Genova, Philippe Esling}
\address{IRCAM - Sorbonne Université, CNRS UMR 9912, 1, place Igor Stravinsky, Paris, France}
\begin{document}
%
\maketitle
\begin{abstract} 
Despite significant advances in deep models for music generation, the use of these techniques remains restricted to expert users. Before being democratized among musicians, generative models must first provide \textit{expressive control} over the generation, as this conditions the integration of deep generative models in creative workflows. In this paper, we tackle this issue by introducing a deep generative audio model providing expressive and continuous descriptor-based control, while remaining lightweight enough to be embedded in a hardware synthesizer. We enforce the controllability of real-time generation by explicitly removing salient musical features in the latent space using an adversarial confusion criterion. User-specified features are then reintroduced as additional conditioning information, allowing for \textit{continuous} control of the generation, akin to a synthesizer knob. We assess the performance of our method on a wide variety of sounds including instrumental, percussive and speech recordings while providing both \textit{timbre} and \textit{attributes} transfer, allowing new ways of generating sounds. 

\def\thefootnote{*}\footnotetext{These authors contributed equally to this work}


\end{abstract}
\section{Introduction}
\label{sec:intro}

In recent years, deep generative models have offered exciting new ways to synthesize sound and accomplished impressive results in generating high-quality audio samples. Early works on auto-regressive (AR) models successfully achieved high-quality raw waveform synthesis \cite{mehri2016samplernn}, but at the cost of expensive computation \cite{esling2020diet}. Subsequent approaches have leveraged a time-frequency representation of the signals and reduce significantly the inference cost \cite{kumar2019melgan}, but remain hard to control. These methods mainly rely on Generative Adversarial Networks (GAN) which produces highly realistic samples \cite{goodfellow2014generative} or Variational AutoEncoders (VAE) \cite{kingmavae} which provides a latent representation that captures high-level signal features.

The RAVE model \cite{rave} leverages both VAE representation abilities and GAN training to achieve high-quality waveform synthesis in real-time on a standard laptop CPU. However, controlling the generation process over non-differentiable attributes remains a challenging task. Prior works tried to enhance generative models with control attributes \cite{perez2018film, yang2021multi}, notably the Conditional VAE (C-VAE) \cite{sohn2015learning}, which adds an attribute vector as input to the encoder and decoder to condition the generation \cite{sohn2015learning}. However, it can lead to poor control abilities as the model could ignore the conditioning and sample directly from the latent variables. To address this, Fader Networks \cite{lample2017fader} force the latent representation to be invariant to given target attributes through adversarial learning. This work has been extended for VAEs to control real-valued attributes instead of binary tags for symbolic generation \cite{kawai2020attributes}. 

Yet, providing \textit{high-level, continuous and time-varying} controls with perceptually meaningful parameters over the raw audio waveform remains an open challenge. In this paper, we propose to tackle this issue by applying the \textit{Fader Networks} approach on top of the state-of-art deep audio model RAVE \cite{rave} in order to provide intuitive controls for sound synthesis. After explicitly disentangling salient musical features in the RAVE latent space by relying on the \textit{Fader} adversarial confusion criterion, we reintroduce them as additional inputs to the RAVE decoder to condition the generation, allowing for continuous time-varying control akin to a synthesizer knob. We show that our model provides more accurate control compared to our baselines on various speech and instrumental datasets. We evaluate the impact of control on the generation quality using various quality metrics. Unlike prior works, our method orthogonalizes descriptor-based control attributes from the latent representation. Hence, it provides independent priors on both the latent representation and the control attributes, allowing to separately (or jointly) perform \textit{timbre transfer} and \textit{attribute transfer}. Hence, the user can choose any set of audio descriptors to condition the generation process. Our approach remains lightweight and can be embedded into highly-constrained hardware synthesizers\footnote{All of our source code and experiments are available on a supplementary webpage: \url{https://github.com/neurorave/neurorave}}.


\begin{figure*}[h!]
 \centerline{
 \includegraphics[width=1.9\columnwidth]{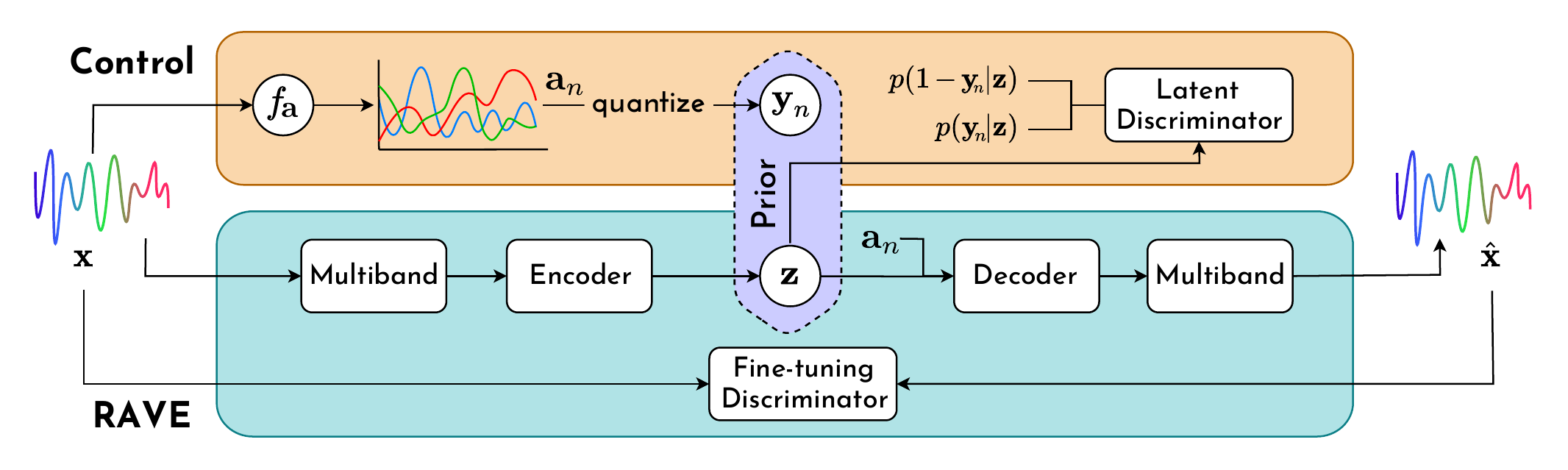}}
 \vspace{-5pt}
 \caption{Overall workflow of our proposed method combining RAVE, Fader network and joint prior training.}
 \vspace{-5pt}
 \label{fig:faderrave}
\end{figure*}

\section{Proposed model}\label{sec:soa}

We aim to provide \textit{continuous attributes control} on any type of audio descriptor for \textit{raw waveform generation}. To do so, we propose to regularize the latent space of a generative model with an adversarial criterion and condition the generation with this given set of continuous controls. This overall architecture is depicted in Figure \ref{fig:faderrave}. 




\paragraph*{Generative models.} We base our work on RAVE \cite{rave} as it allows fast and high quality audio waveform synthesis (with sampling rates up to 48kHz) on any type of signal while being usable with real-time constraints on a laptop CPU. To do so, it leverages a multi-band decomposition of the raw waveform with a \textit{Pseudo Quadrature Mirror Filter bank} (PQMF) \cite{yu2019durian, yang2021multi}, which allows to decrease the temporal dimensionality of the data. Similarly to the RAVE \cite{rave} model (depicted in blue on Figure \ref{fig:faderrave}), the training is split between a \textit{representation learning} stage and an \textit{adversarial fine-tuning} stage. 

The first stage aims to learn relevant features for the latent space by optimizing a multiscale spectral distance \cite{ddsp}. The \textit{encoder} and \textit{decoder} are trained to minimize the loss $\mathcal{L}_{vae}$ derived from the ELBO \cite{kingmavae}. Hence, for a given signal $\mathbf{x} \in \mathbb{R}^{n}$ (where $n$ is the initial discrete time length), we train a VAE to reconstruct this input signal by learning an informative and compact latent representation. This representation keeps the temporal dimension as it produces a matrix $\mathbf{z} \in \mathbb{R}^{d \times m}$, where $d < n$ is the number of latent dimensions and $m\in\mathbb{N}$ is the compressed discrete time length. This dimension depends on the sampling rate and compression ratio applied by the encoder. Therefore, the embedded representation can be seen as a temporal trajectory in a $d$-dimensional space.  Once  the latent representation is learned, the \textit{encoder} is frozen to perform the \textit{adversarial fine-tuning} stage.

During the second stage, the latent space is seen as a base GAN distribution and the \textit{decoder} learns to generate more realistic signals by relying on a \textit{discriminator} $D$ optimizing the Hinge loss $\mathcal{L}_g$ \cite{lim2017geometric}.
To ensure that the synthesized signal $\mathbf{\hat{x}}$ does not diverge too much from the ground truth $\mathbf{x}$, the model keeps minimizing the multiscale spectral distance, but also adds the feature matching loss $\mathcal{L}_{FM}$ proposed in \cite{kumar2019melgan}, which minimizes the $L_1$ distance between the \textit{discriminator feature maps} of real and synthesized audio. Hence, the full objective loss for the \textit{decoder} becomes
\begin{equation}
\begin{split}
    \label{eqn:lossdec}
    \mathcal{L}_{dec} = \mathcal{L}_{vae} + \mathcal{L}_{g} + \mathcal{L}_{FM}.
\end{split}
\end{equation}


\paragraph*{Control.} Within the image field, one of the first approach of control \cite{sohn2015learning} adds the features information as conditioning input to a VAE. GAN models also incorporated attributes control, such as in StyleGAN \cite{azadi2018multi} which splits high-level features from stochastic variation, or AttGans \cite{8718508} which includes an attribute classification constraint to the generated image. Within the audio generation domain, DDSP \cite{ddsp} relies on explicit control signals, but only offers pitch and loudness modifications over the generation.

Fader Networks \cite{lample2017fader} is a particularly interesting approach as it provides explicit control on any set of desired attributes. It is achieved by applying an adversarial discriminator in the latent space, forcing the learning of invariant representations with respect to the varying attributes. This discriminator is trained to predict the real attributes from the latent representation, while the encoder is forced to bypass all features information in order to prevent the discriminator from achieving its goal. This implies the adversarial losses
\begin{equation}
\begin{split}
    \label{eqn:latdis_loss1}
    \mathcal{L}_{dis}(\theta_{dis} | \theta_{enc}) = - \mathbb{E}_{p_{\textit{enc}}(\mathbf{z} | \mathbf{x})}[\text{log}(p_{\textit{dis}}(\mathbf{y} | \mathbf{z}))] \\
    \mathcal{L}_{dis}(\theta_{enc} | \theta_{dis}) = - \mathbb{E}_{p_{\textit{enc}}(\mathbf{z} | \mathbf{x})}[\text{log}(p_{\textit{dis}}(1-\mathbf{y} | \mathbf{z}))],
\end{split}
\end{equation}
where $\theta_{enc}$ denotes the encoder parameters, $\theta_{dis}$ those of the discriminator and $p_{\textit{dis}}(\mathbf{y} | \mathbf{z})$ represents the (discriminator) probability of an attribute vector $\mathbf{y}$ given a latent representation $\mathbf{z}$. The drawback of this method is that it only considers binary and static attributes, which are insufficient in the context of audio generation control. However, \cite{kawai2020attributes} relies on an extension of Faders to provide control over symbolic musical attributes. Although their method allows for real-valued parameters instead of binary attributes, they still do not address the problem of continuous temporal control. Hence, we propose to extend this method to provide both continuous high-level attributes control and raw audio generation.

\paragraph*{Our approach.} The \textit{representation learning} stage of training perfectly fits with the first objective of Fader networks, which is to re-organize an encoded representation. Once the model converges, the \textit{encoder} and \textit{fader discriminator} are frozen and the second \textit{adversarial fine-tuning} stage begins in order to improve the quality of the generation. In order to introduce control (depicted in orange in Figure \ref{fig:faderrave}), we add an adversarial \textit{discriminator} to the latent space during the first stage similarly to \textit{Fader Networks} \cite{lample2017fader}. The \textit{encoder} is enforced to learn a representation invariant to the chosen control attributes $\mathbf{y}_{n} \in \mathbb{R}$, $\forall n \in  [1 , N]$, where $N$ is the number of considered attributes. However, contrarily to previous approaches, we perform on-the-fly attribute computation from input $\mathbf{x}$, such that $\mathbf{a}_{n}=f_{a}(\mathbf{x})$, with $f_a$ being the descriptor computation function. Then, the \textit{decoder} receives the audio descriptor $\mathbf{a}_{n}$, which is resampled to match the temporal dimension, as additional information to the latent matrix $\mathbf{z}$. 

In order to constrain the \textit{decoder} to use the attribute information to reconstruct the signal, the \textit{latent discriminator} gives feedback to the \textit{encoder} by optimizing the adversarial criterion, which forces the \textit{encoder} to remove any attribute information from the latent space. The attributes $\mathbf{y}_n$ used for this training are obtained from the conversion of continuous attributes $\mathbf{a}_{n}$ into discrete categorical quantities by performing a quantization. It is achieved by applying an RMS threshold to remove silent parts of the sound and then sorting these values to obtain bins with equal density.  The final training loss is defined by the combination of Eq.~\ref{eqn:lossdec} and Eq.~\ref{eqn:latdis_loss1}
\begin{equation}
\begin{split}
    \label{eqn:losstotal}
    \mathcal{L} & = \mathcal{L}_{vae} + \lambda \cdot \mathcal{L}_{dis}(\theta_{enc} | \theta_{dis}),
\end{split}
\end{equation}
where the $\lambda$ hyperparameter controls the trade-off between the ELBO and the invariance of the latent representation.

\section{Experiments}\label{sec:exp}

\paragraph*{Audio descriptors.} Descriptors provide quantitative measures on the properties of a given sound \cite{peeters2011timbre}. To perform a direct \textit{descriptor-based} synthesis, we selected a subset of \textit{local descriptors} (providing a value for each frame of signal), which are also complex to differentiate, while being linked to perceptual properties. We only retain descriptors that are pairwise independent to successfully disentangle the selected attributes from the latent representation. In addition to spectral descriptors (\textit{RMS} perceptually similar to \textit{loudness}, \textit{Centroid} to \textit{brightness}, and \textit{Bandwidth} to \textit{richness}), we rely on more advanced timbral features provided by \textit{Audio Commons}\footnote{  \url{http://github.com/AudioCommons/timbral\_models}}: \textit{Sharpness} and \textit{Boominess}, which we adapt to be computed frame-wise rather than globally.

\vspace{-12pt}
\paragraph*{Datasets.} We experiment on a variety of dataset divided between train (80\%), validation (10\%) and test (10\%) splits.
\vspace{-12pt}
\paragraph*{\textit{Instrumental (NSynth \cite{engel2017neural})}} is a collection of instruments with 4-seconds monophonic sounds sampled at 16kHz. We use only \textit{string} family, which corresponds to 20594 samples.

\vspace{-12pt}
\paragraph*{\textit{Percussive (Darbuka)}} is an internal dataset composed of about 2 hours of raw recordings of darbuka sampled at 48kHz.
\vspace{-20pt}
\paragraph*{\textit{Speech (SC09)}} is part of the \textit{Speech Commands} \cite{warden2018speech} set containing the spoken digits from multiple speakers in various recording conditions sampled at 16kHz (23666 samples).

\begin{table*}[h!]
 \begin{center}

\begin{tabular}{c|ccc|cc|c||ccc|cc|c}
\hline 
 & \multicolumn{3}{c|}{\textbf{Reconstruction}} & \multicolumn{2}{c|}{\textbf{Control}} & \textbf{Cycle} & \multicolumn{3}{c|}{\textbf{Reconstruction}} & \multicolumn{2}{c|}{\textbf{Control}} & \textbf{Cycle}\tabularnewline
\hline 
\hline 
 & \emph{JND} & \emph{Mel} & \emph{mSTFT} & \emph{Corr.} & \emph{L1} & \textit{JND} & \emph{JND} & \emph{Mel} & \emph{mSTFT} & \emph{Corr.} & \emph{L1} & \textit{JND}\tabularnewline
\hline 
RAVE & 0.264 & 15.097 & 7.754 & - & - & - & 0.264 & 15.097 & 7.754 & - & - & -\tabularnewline
\hline 
C-RAVE & \textbf{0.225} & \textbf{12.567} & 5.842 & 0.890 & 0.135 & 0.603 & 0.208 & \textbf{12.236} & \textbf{4.584} & 0.425 & 0.361 & 0.752\tabularnewline
\hline 
F-RAVE & 0.238 & 13.876 & \textbf{5.394} & \textbf{0.917} & \textbf{0.112} & \textbf{0.581} & \textbf{0.187} & 13.050 & 4.681 & \textbf{0.445} & \textbf{0.357} & \textbf{0.776}\tabularnewline
\hline 
\end{tabular}
\end{center}
 \vspace{-10pt}
 \caption{Comparison of various models on the mono-attribute \textit{rms} training (left), and the complete multi-attribute training (right). We compare these on their quality of reconstruction, ability to control attributes and cycle consistency.}
 \label{tab:full_models_results}
\end{table*}

\paragraph*{Implementation details} All experiments are conducted using the ADAM optimizer \cite{kingma2014adam} with $\beta_1 = 0.5$, $\beta_2 = 0.9$, a batch size of 8 and a learning rate of $0.0001$ as suggested in \cite{rave}. The weights are initialized from a Normal distribution $\mathcal{N}(0, 0.02)$ and activations are LeakyReLU with a leak of $0.2$. We implemented the same architecture as the original RAVE paper for the \textit{encoder}, \textit{decoder} and \textit{discriminator} with an $8$-band PQMF decomposition. 
For the \textit{latent discriminator} architecture, we implemented a series of three 1-d convolutional blocks per attributes along the latent dimensions $d$, in order to preserve the temporal dimension $m$ of the latent space. Each block is composed of convolution, BatchNorm and LeakyReLU, with an additional softmax activation on the last convolution layer.
We train the \textit{representation learning} stage with \textit{Fader} discriminator for $1000k$ steps. The $\beta$ warmup from $0$ to $0.1$ and fader $\lambda$ warmup from $0$ to $0.5$, start respectively at step $5k$ and $15k$, and end at step $10k$ and $30k$.

\paragraph*{Baselines.}  To evaluate our method, we implement the \textit{RAVE} model as a baseline for reconstruction accuracy. Then we add to it a conditioning mechanism (concatenating to the latent vector the repeated temporal attributes as extra channels) leading to a conditional-RAVE (\textit{C-RAVE}) model. Finally, we compare them to Fader-Rave (\textit{F-RAVE}) described in Sec~\ref{sec:soa}.

\paragraph*{Evaluation.} For the evaluation of the \textit{reconstruction}, we use the multiscale-STFT (\textit{mSTFT}). As this is also our training criterion, we rely on two independent spectral metrics: the $L_{1}$ distance between Mel-spectrograms (\textit{Mel}) and the \textit{Just Noticeable Difference} (\textit{JND}) score \cite{manocha2020differentiable}. Regarding the efficiency of \textit{attributes control}, we compute the correlation between changed control attributes and resulting output attributes through the Spearman rank-order correlation coefficient. We also compute the \textit{$L_1$} distance to evaluate quantitatively how the generation reflects the control. Additionally, we perform a \textit{cycle consistency} distance to ensure the model is able to  reproduce the original audio when reverting the attributes of a transformed signal to their original values.
 
\section{Results}\label{sec:results}    

\paragraph*{Models quality and control.} We compare the quality of reconstruction for inputs from the test set for \textit{RAVE}, \textit{C-RAVE} and \textit{F-RAVE}. Then, we evaluate the control behaviour by changing the attributes of the input sound with those of out-of-distribution examples (e.g. we switch the attributes of the violin for those of the darbuka). We compute this for mono-attribute training (swapping only the \textit{RMS}) or for multi-attribute training (swapping all attributes) cases. We detail our results obtained on \textit{Nsynth} \cite{engel2017neural} in Table~\ref{tab:full_models_results}.

Interestingly, introducing conditioning (C-RAVE) seems to improve the overall reconstruction quality, in both the mono- (left) and multi- (right) cases. This could be explained by the fact that this extraneous information simplifies the work of the decoder, by providing structured information on the generation. Although the C-RAVE model is able to adequately control the single \textit{RMS} attribute case, it loses in control quality for the more complicated multi-attribute case (although it still improves reconstruction). Oppositely, F-RAVE provides stronger correlation in mono-attribute RMS control, while maintaining equivalent audio quality. It seems also more resilient to the multi-attribute changes, providing the strongest correlation in control and the lowest reconstruction error. This implies that the model could be trained once for a whole set of descriptors, while still maintaining control quality. This is also reflected by the cycle consistency, which appears more coherent than the C-RAVE model.

\paragraph*{Multiple control attributes.} We also analyze the control quality by training a separate model for each of the 4 descriptors, and a model for all descriptors at once (termed C-RAVE (m.) and F-RAVE (m.)). We analyze the correlation between target and output attributes when changing a single descriptor, or when changing random sets of 2, 3, or 4 attributes at once. The results obtained with \textit{Nsynth} \cite{engel2017neural} are displayed in  Table~\ref{tab:results_control_mono_multi}.

\begin{table}[h!]
 \begin{center}
\begin{tabular}{c|cccc}
\hline 
\textbf{Mono-attribute} & \emph{RMS} & \emph{Centro} & \emph{Sharp.} & \emph{Boom.}\tabularnewline
\hline 
C-RAVE & 0.890 & \textbf{0.310} & \textbf{0.796} & 0.760\tabularnewline
F-RAVE & \textbf{0.917} & 0.282 & 0.760 & \textbf{0.773}\tabularnewline
\hline 
\hline 
\textbf{Multi-attribute} & \textit{1} & \textit{2} & \textit{3} & \textit{4}\tabularnewline
\hline 
C-RAVE (m.) & 0.602 & 0.524 & 0.463 & 0.425\tabularnewline
F-RAVE (m.) & \textbf{0.616} & \textbf{0.539} & \textbf{0.478} & \textbf{0.445}\tabularnewline
\hline 
\end{tabular}

\end{center}
 \vspace{-10pt}
 \caption{Comparison of the control provided by various models on different descriptors for \textit{mono-attribute} (top) or \textit{multi-attribute} (bottom) setups.}
 \label{tab:results_control_mono_multi}
\end{table}

As we can see (up), models that are trained on single attributes provide a stronger control over that descriptor. In these, it seems that \textit{RMS}, the most salient descriptor is the easiest to control, while more complex concepts (\textit{Sharpness}, \textit{Boominess}, \textit{Centroid}) are harder to grasp for the control models. As can be expected, relying on a multi-attribute model rather than specifically-trained models impact the quality of control, especially for simple conditioning (C-RAVE). However, \textit{F-RAVE} model seems to be less affected by this property. Similarly, when increasingly changing a larger number of attributes (bottom), the \textit{C-RAVE (m.)} steadily degrades in terms of control quality, while the \textit{F-RAVE (m.)} model maintains a higher quality of control, even when switching all attributes of a sound.

\paragraph*{Comparing various datasets.} We finally evaluate how our proposed F-RAVE can be used on a wide diversity of sounds in the multi-attribute setup. We display the reconstruction (\textit{Rec.}) and control (\textit{Ctr.}) results in Table~\ref{tab:results_datasets}.

\begin{table}[h!]
\begin{center}
\begin{tabular}{c|cc|cc|cc}
\cline{2-7} \cline{3-7} \cline{4-7} \cline{5-7} \cline{6-7} \cline{7-7} 
\multicolumn{1}{c}{} & \multicolumn{2}{c|}{\textbf{NSynth}} & \multicolumn{2}{c|}{\textbf{Darbouka}} & \multicolumn{2}{c}{\textbf{SC09}}\tabularnewline
\hline 
\hline 
 & \emph{Rec.} & \emph{Ctr.} & \emph{Rec.} & \emph{Ctr.} & \emph{Rec.} & \emph{Ctr.}\tabularnewline
\hline 
C-RAVE & \textbf{4.58} & 0.42 & 5.31 & 0.62 & 6.41 & \textbf{0.35}\tabularnewline
\hline 
F-RAVE & 4.68 & \textbf{0.45} & \textbf{5.14} & \textbf{0.61} & \textbf{6.33} & 0.34\tabularnewline
\hline 
\end{tabular}
\end{center}
 \vspace{-10pt}
 \caption{Results on the \textit{instrumental} (NSynth), \textit{percussive} (Darbouka) and \textit{speech} (SC09) datasets.}
 \label{tab:results_datasets}
\end{table}

RAVE appears versatile enough to maintain a high quality on any type of sounds, and confirm that the reconstruction quality stands even under multiple attribute conditioning. Regarding other models, the trends from the previous results seem to be maintained across different datasets, with a slight advantage for the F-RAVE model. The percussive dataset appears to be the easiest to control, which could be explained by the larger variance in the descriptor values for this set. 

\section{Conclusion}\label{sec:conclusion}

In this paper, we combined \textit{Fader Networks} and the recent \textit{RAVE} model in order to achieve continuous descriptor-based control on real-time deep audio synthesis. We showed that our method outperforms previous proposals in both quantitative and qualitative analyses. By orthogonalizing the continuous time-varying attributes from the latent representation, our approach provides independent priors which allows to separately perform both \textit{timbre transfer} and \textit{attribute transfer} enabling new creative prospects. Hence, the user can choose a set of descriptors to condition the generation process and create a large variety of sounds. Altogether, we hope this expressive control approach will benefit both experts and non-experts musicians and provide new means of exploring audio synthesis while promoting co-creative musical applications.

\newpage
\bibliographystyle{IEEEbib}
\bibliography{strings,refs}

\begin{thebibliography}{10}

\bibitem{mehri2016samplernn}
Soroush Mehri, Kundan Kumar, Ishaan Gulrajani, Rithesh Kumar, Shubham Jain,
  Jose Sotelo, Aaron Courville, and Yoshua Bengio,
\newblock ``Samplernn: An unconditional end-to-end neural audio generation
  model,''
\newblock {\em arXiv preprint arXiv:1612.07837}, 2016.

\bibitem{esling2020diet}
Philippe Esling, Ninon Devis, Adrien Bitton, Antoine Caillon, Constance Douwes,
  et~al.,
\newblock ``Diet deep generative audio models with structured lottery,''
\newblock {\em arXiv preprint arXiv:2007.16170}, 2020.

\bibitem{kumar2019melgan}
Kundan Kumar, Rithesh Kumar, Thibault de~Boissiere, Lucas Gestin, Wei~Zhen
  Teoh, Jose Sotelo, Alexandre de~Br{\'e}bisson, Yoshua Bengio, and Aaron~C
  Courville,
\newblock ``Melgan: Generative adversarial networks for conditional waveform
  synthesis,''
\newblock {\em Advances in neural information processing systems}, vol. 32,
  2019.

\bibitem{goodfellow2014generative}
Ian Goodfellow, Jean Pouget-Abadie, Mehdi Mirza, Bing Xu, David Warde-Farley,
  Sherjil Ozair, Aaron Courville, and Yoshua Bengio,
\newblock ``Generative adversarial nets,''
\newblock {\em Advances in neural information processing systems}, vol. 27,
  2014.

\bibitem{kingmavae}
Diederik~P Kingma and Max Welling,
\newblock ``Auto-encoding variational bayes,''
\newblock {\em arXiv preprint arXiv:1312.6114}, 2013.

\bibitem{rave}
Antoine Caillon and Philippe Esling,
\newblock ``{RAVE:} {A} variational autoencoder for fast and high-quality
  neural audio synthesis,''
\newblock {\em CoRR}, vol. abs/2111.05011, 2021.

\bibitem{perez2018film}
Ethan Perez, Florian Strub, Harm De~Vries, Vincent Dumoulin, and Aaron
  Courville,
\newblock ``Film: Visual reasoning with a general conditioning layer,''
\newblock in {\em Proceedings of the AAAI Conference on Artificial
  Intelligence}, 2018, vol.~32.

\bibitem{yang2021multi}
Geng Yang, Shan Yang, Kai Liu, Peng Fang, Wei Chen, and Lei Xie,
\newblock ``Multi-band melgan: Faster waveform generation for high-quality
  text-to-speech,''
\newblock in {\em 2021 IEEE Spoken Language Technology Workshop (SLT)}. IEEE,
  2021, pp. 492--498.

\bibitem{sohn2015learning}
Kihyuk Sohn, Honglak Lee, and Xinchen Yan,
\newblock ``Learning structured output representation using deep conditional
  generative models,''
\newblock {\em Advances in neural information processing systems}, vol. 28,
  2015.

\bibitem{lample2017fader}
Guillaume Lample, Neil Zeghidour, Nicolas Usunier, Antoine Bordes, Ludovic
  Denoyer, and Marc'Aurelio Ranzato,
\newblock ``Fader networks: Manipulating images by sliding attributes,''
\newblock {\em Advances in neural information processing systems}, vol. 30,
  2017.

\bibitem{kawai2020attributes}
Lisa Kawai, Philippe Esling, and Tatsuya Harada,
\newblock ``Attributes-aware deep music transformation,''
\newblock in {\em Proceedings of the 21st international society for music
  information retrieval conference, ismir}, 2020.

\bibitem{yu2019durian}
Chengzhu Yu, Heng Lu, Na~Hu, Meng Yu, Chao Weng, Kun Xu, Peng Liu, Deyi Tuo,
  Shiyin Kang, Guangzhi Lei, et~al.,
\newblock ``Durian: Duration informed attention network for multimodal
  synthesis,''
\newblock {\em arXiv preprint arXiv:1909.01700}, 2019.

\bibitem{ddsp}
Jesse Engel, Lamtharn Hantrakul, Chenjie Gu, and Adam Roberts,
\newblock ``Ddsp: Differentiable digital signal processing,''
\newblock {\em arXiv preprint arXiv:2001.04643}, 2020.

\bibitem{lim2017geometric}
Jae~Hyun Lim and Jong~Chul Ye,
\newblock ``Geometric gan,''
\newblock {\em arXiv preprint arXiv:1705.02894}, 2017.

\bibitem{azadi2018multi}
Samaneh Azadi, Matthew Fisher, Vladimir~G Kim, Zhaowen Wang, Eli Shechtman, and
  Trevor Darrell,
\newblock ``Multi-content gan for few-shot font style transfer,''
\newblock in {\em Proceedings of the IEEE conference on computer vision and
  pattern recognition}, 2018, pp. 7564--7573.

\bibitem{8718508}
Zhenliang He, Wangmeng Zuo, Meina Kan, Shiguang Shan, and Xilin Chen,
\newblock ``Attgan: Facial attribute editing by only changing what you want,''
\newblock {\em IEEE Transactions on Image Processing}, vol. 28, no. 11, pp.
  5464--5478, 2019.

\bibitem{peeters2011timbre}
Geoffroy Peeters, Bruno~L Giordano, Patrick Susini, Nicolas Misdariis, and
  Stephen McAdams,
\newblock ``The timbre toolbox: Extracting audio descriptors from musical
  signals,''
\newblock {\em The Journal of the Acoustical Society of America}, vol. 130, no.
  5, pp. 2902--2916, 2011.

\bibitem{engel2017neural}
Jesse Engel, Cinjon Resnick, Adam Roberts, Sander Dieleman, Mohammad Norouzi,
  Douglas Eck, and Karen Simonyan,
\newblock ``Neural audio synthesis of musical notes with wavenet
  autoencoders,''
\newblock in {\em International Conference on Machine Learning}. PMLR, 2017,
  pp. 1068--1077.

\bibitem{warden2018speech}
Pete Warden,
\newblock ``Speech commands: A dataset for limited-vocabulary speech
  recognition,''
\newblock {\em arXiv preprint arXiv:1804.03209}, 2018.

\bibitem{kingma2014adam}
Diederik~P Kingma and Jimmy Ba,
\newblock ``Adam: A method for stochastic optimization,''
\newblock {\em arXiv preprint arXiv:1412.6980}, 2014.

\bibitem{manocha2020differentiable}
Pranay Manocha, Adam Finkelstein, Richard Zhang, Nicholas~J Bryan, Gautham~J
  Mysore, and Zeyu Jin,
\newblock ``A differentiable perceptual audio metric learned from just
  noticeable differences,''
\newblock {\em arXiv preprint arXiv:2001.04460}, 2020.

\end{thebibliography}

\end{document}